\documentclass[twocolumn,amsmath,amssymb]{revtex4}

\usepackage{multirow}
\usepackage{graphicx}
\begin{document}

\title{Identifying optimal targets of network attack by belief propagation 
  \footnote{Accepted for publication in the journal \emph{Physical Review E} 
    (http://pre.aps.org/). Figure 3 contains more information than that of
    the accepted PRE paper.}}

\author{
  Salomon Mugisha$^{1,2}$ and Hai-Jun Zhou$^{1,2}$\footnote{Corresponding
    author. Email: {\tt zhouhj@itp.ac.cn}}
}

\affiliation{
  $^1$Key Laboratory of Theoretical Physics, Institute of
  Theoretical Physics, Chinese Academy of Sciences,
  Zhong-Guan-Cun East Road 55,
  Beijing 100190, China \\
  $^2$School of Physical Sciences, University of Chinese Academy of Sciences,
  Beijing 100049, China
}

\date{March 18, 2016 (1st version); March 29, 2016 (2nd version); 
  June 28, 2016 (final version)}

\begin{abstract}
  For a network formed by nodes and undirected links between pairs of nodes,
  the network optimal attack problem aims at deleting a minimum number of
  target nodes to break the network down into many small components. This
  problem is intrinsically related to the feedback vertex set problem that was
  successfully tackled by spin glass theory and an associated belief
  propagation-guided decimation (BPD) algorithm
  [H.-J. Zhou, Eur.~Phys.~J.~B {\bf 86} (2013) 455]. In the present work we
  apply the BPD alrogithm (which has approximately linear time complexity) to
  the network optimal attack problem, and demonstrate that it has much better
  performance than a recently proposed Collective Information algorithm 
  [F. Morone and H. A. Makse, Nature {\bf 524} (2015) 63--68] for different
  types of random networks and real-world network instances. The BPD-guided
  attack scheme often induces an abrupt collapse of the whole network, which
  may make it very difficult to defend.
\end{abstract}

\maketitle

\section{Introduction}
\label{sec:intro}

Consider a network or graph $G$ formed by $N$ nodes and $M$ undirected links
between these nodes, how to delete a minimum number of nodes (the optimal
targets of attack) to break the network down into many disconnected small
components? This optimization problem is one of the fundamental structural
problems in network science \cite{He-Liu-Wang-2009,Albert-Barabasi-2002}, and
it has very wide practical applications, especially in protection of network
structure \cite{Albert-Jeong-Barabasi-2000,Callaway-etal-2000,Cohen-etal-2001}
and in surveillance and control of various network dynamical processes such as
the transmission of infective disease
\cite{PastorSatorras-Vespignani-2001,Altarelli-etal-2014,Guggiola-Semerjian-2015}.
Besides their structural importance, the optimal target nodes of network
attack also play significant roles in network information diffusion. Many of
these nodes are influential spreaders of information and are the key objects in
viral marketing and network advertisement
\cite{Richardson-Domingos-2002,Kempe-etal-2015,Altarelli-Braunstein-DallAsta-Zecchina-2013}.

The breakdown of a network's giant connected component is the collective effect
caused by a set $S$ of nodes. There are extremely many candidate solutions for
the network attack problem, and minimizing the size of such a set $S$ is 
an intrinsically difficult combinatorial optimization issue. This problem
belongs to the NP-hard (non-deterministic polynomial hard) class of
computational complexity, no one expects it to
be exactly solvable by any polynomial algorithm. So far the network optimal
attack problem has mainly been approached by heuristic methods which select
target nodes based on local metrics such as the node degree (number of attached
links) \cite{Albert-Jeong-Barabasi-2000,Callaway-etal-2000,Cohen-etal-2001} and
the node eigenvector centrality \cite{Bonacich-1987,Li-Zhang-etal-2012}.

For sparse random networks it is well known that the typical length of loops
diverges with the number $N$ of nodes in a linear way, and short loops of
length $L \ll \ln(N)$ are very rare
\cite{Marinari-Monasson-2004,Marinari-etal-2005,Bianconi-Marsili-2005}.
In such networks the small connected components are mostly trees (no loop
inside), while each giant component includes a finite fraction of all the nodes
and an exponential number of long loops. If these long loops are all cut the
giant component will again break into a set of small tree components. For
random network ensembles, therefore, the optimal attack problem is essentially
equivalent to another celebrated global optimization, namely the minimum
feedback vertex set problem \cite{Karp-1972}. A feedback vertex set (FVS) for a
network $G$ is a set of nodes which, if deleted, would break all the loops in
the network and leave behind a forest (that is, a collection of tree
components). In other words, a FVS is a node set that intersects with every
loop of the network, and a minimum FVS is just a node set of smallest size
among all the feedback vertex sets. Because small components of spare random
networks are mostly trees, a minimum FVS is essentially a minimum set of target
nodes for the network attack problem.

Although the minimum FVS problem is also NP-hard, a very convenient mapping of
this optimization problem to a locally constrained spin glass model was
achieved in 2013 \cite{Zhou-2013}. By applying the replica-symmetric mean field
theory of statistical mechanics to this spin glass model, the minimum FVS sizes
and hence also the minimum numbers of targeted attack nodes are quantitatively
estimated for random Erd\"os-Reny\'i (ER)  and random regular (RR) network
ensembles \cite{Zhou-2013}, which are in excellent agreement with rigorously
derived lower bounds \cite{Bau-Wormald-Zhou-2002} and simulated-annealing
results \cite{Qin-Zhou-2014,Galinier-Lemamou-Bouzidi-2013}. Inspired by the
spin glass mean field theory, an efficient minimum-FVS construction algorithm,
belief propagation-guided decimation (BPD), was also introduced in
\cite{Zhou-2013}, which is capable of constructing  close-to-minimum feedback
vertex sets for single random network instances and also for correlated
networks. To solve the optimal attack problem for a network containing a lot of
short loops, the BPD algorithm can be adjusted slightly by allowing the
existence of loops within each small connected component. Such a BPD algorithm
can produce a nearly-minimum set of target nodes to break the giant components.

In 2015, Morone and Makse considered the network optimal attack problem as an
optimal influence problem and introduced an interesting heuristic Collective
Information (CI) algorithm  \cite{Morone-Makse-2015}. These authors called the
optimal targets of network attack as the optimal influencers of the network to
emphasize their importance to information spreading. In the CI algorithm, each
node $i$ is assigned an impact value which counts the number of out-going links
at the surface of a `ball' of radius $\ell$ centered around $i$; and then the
highest-impact nodes are sequentially deleted from the network (and the impact
values of the remaining nodes are updated) until the largest component of the
remaining network becomes sufficiently small. This CI algorithm was tested on
random networks and a set of real-world networks and it was claimed that it
beats existing heuristic algorithms \cite{Morone-Makse-2015}. Morone and Makse
also compared the results obtained by CI and BPD on a single random scale-free
network and they found ``evidence of the best performance of CI''
\cite{Morone-Makse-2015}.

The CI algorithm is local in nature, it considers only the local structure
within distance $\ell$ to each focal node to build the node importance metric.
The claim that such a local-metric algorithm is capable of beating the BPD
algorithm, a distributed message-passing algorithm taking into account the
global loop structure of the network, is indeed quite surprising. Given the
importance of the optimal attack problem in network science, and considering
that only a single network instance was checked in \cite{Morone-Makse-2015},
we believe it will be beneficial to the research community for us to give a
detailed description of the BPD algorithm for the optimal attack problem and
to perform a systematic comparative study on the CI and the BPD algorithm. In
the present paper, after reviewing the most essential building blocks of the CI
and the BPD algorithm, we describe simulation results obtained on three random
network ensembles (random ER and RR networks, whose structures are homogeneous;
and random scale-free networks, whose structures are heterogeneous), and a set
of real-world network instances (whose structures are heterogeneous and highly
correlated, and there are an abundant number of short loops inside).

Our extensive simulation results convincingly demonstrate that the BPD
algorithm offers qualitatively superior solutions to the network optimal attack
problem for random and real-world networks. Our data reveal that, both for
random and for real-world networks, the solutions constructed by the CI
algorithm are far from being optimal. For example, to break an internet network
instance (IntNet2 of Table~\ref{tab:realnet}, with $N \approx 1.7 \times 10^6$
nodes) following the recipe offered by CI one would have to attack
$\approx 1.4 \times 10^5$ nodes simultaneously, but actually the job can be
finished by attacking only $\approx 7.3 \times 10^4$ nodes if instead the
recommendations of the BPD algorithm are adopted. For sparse networks the
running time of the BPD algorithm scales almost linearly with the number $N$ of
nodes in the network, so it is ideally suitable for treating network instances
of extreme sizes.

Let us close this introductory section by pointing out a potential challenge
that network defense practitioners might have to consider in the near future. 
Imagine that certain group of antisocial agents (e.g., terrorists) plans to 
carry out an intentional distributed network attack by destroying a small set
of target nodes specified by the BPD algorithm or other loop-focused global
algorithms. Under such a BPD-guided distributed attack, our example results
of Fig.~\ref{fig:ERattack} (solid line) and Fig.~\ref{fig:realnet} suggest that
the network remains to be globally intact and connected before it undergoes a
sudden and abrupt collapse. For the defense side, such a `no serious warning'
situation might make it very difficult to distinguish between intentional
attacks and random localized failures and to carry out timely reactions. We
leave this issue of theoretical and practical importance to further serious
investigations.

\section{A brief review on CI and BPD}
\label{sec:methods}

As we already introduced, the goal of the network optimal attack problem is to
construct a minimum  node set $S$ for an input network $G$ so that the
sub-network induced by all the  nodes not in $S$ has no connected component of
relative size exceeding certain small threshold  $\theta$ (e.g., $\theta=0.01$
or even smaller). The CI algorithm of \cite{Morone-Makse-2015} and the BPD
algorithm of \cite{Zhou-2013} are two heuristic solvers for this NP-hard
optimization problem. For pedagogical reasons we summarize in this section the
main algorithmic steps of these two solvers. We do not delve into the
underlying statistical physical ideas and concepts but encourage the reader to
consult the original references.

Starting from the input network $G$ with $N$ nodes and $M$ links,  both the CI
and the BPD algorithm kick nodes out of the network in a sequential manner. Let
us denote by $G(t)$ the remaining network at time $t$ of the deletion process,
and denote by $d_i(t)$ the degree (number of neighboring nodes) of a node $i$
in $G(t)$. At the initial time $t=0$ all the nodes are present so $G(0)$ is
identical to $G$, and $d_i(0)=d_i$ with $d_i$ being the degree of node $i$ in
$G$.

\subsection{The Collective Influence algorithm}
\label{subsec:CIA}

At each time point $t$ the collective influence strength, $CI_{\ell}(i; t)$, of 
a node $i \in G(t)$ is computed as
\begin{equation}
  \label{eq:CI}
  CI_{\ell}(i; t)= \bigl[ d_i(t)-1\bigr]
  \sum\limits_{j\in \partial {\textrm{Ball}(i,\ell;\, t)}} \bigl[d_{j}(t)-1\bigr]\; ,
\end{equation}
where $\partial \textrm{Ball}(i, \ell;\, t)$ denotes the set formed by all the
nodes of $G(t)$ that are at distance $\ell$ to node $i$
\cite{Morone-Makse-2015}. The integer $\ell$ is an adjustable parameter of the
CI algorithm. The CI strength gives a heuristic measure of a node's information
spreading power. It is a product of two terms. The first term,
$\bigl(d_i(t)-1\bigr)$, is node $i$'s direct capacity of information
transmission; the second term sums over the information transmission capacity
$\bigl(d_j(t)-1\bigr)$ of all the nodes $j$ at a distance $\ell$, it can be
understood as node $i$'s capacity of information broadcasting.

After the CI strengths of all the nodes in network $G(t)$ are updated using
Eq.~(\ref{eq:CI}), a node which has the highest CI strength is deleted along
with all its attached links; then the time increases to
$t\leftarrow t+ \frac{1}{N}$, and the CI strength of the remaining nodes are
again updated. This iteration process continues until the largest connected 
component of the remaining network becomes very small.

\begin{figure}
  \begin{center}
    \includegraphics[angle=270,width=0.45\textwidth]{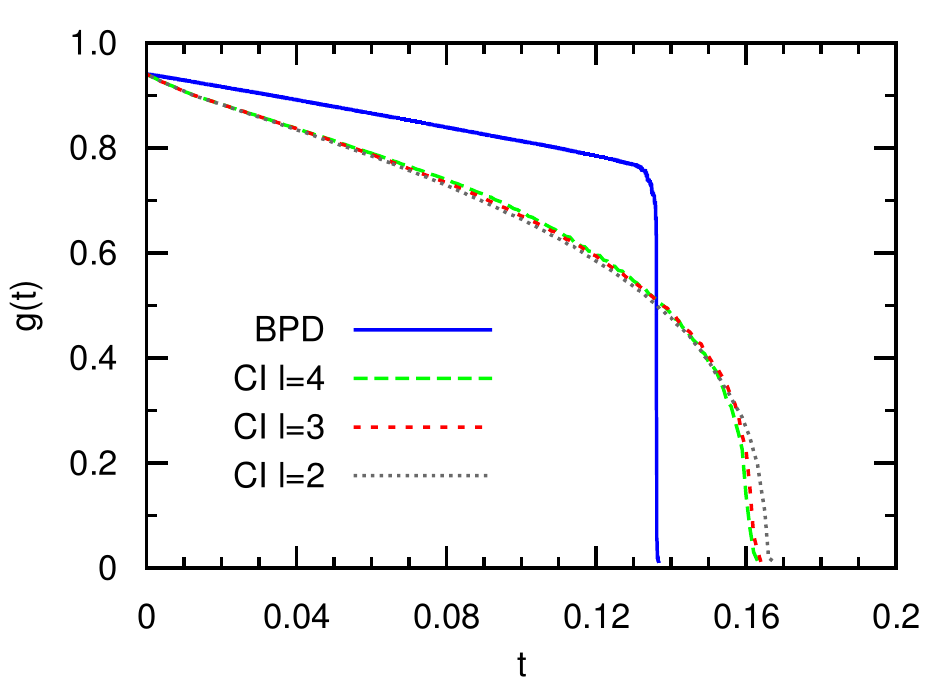}
  \end{center}
  \caption{\label{fig:ERattack}
    The relative size $g(t)$ of the largest connected component as a function of
    algorithmic time $t$, for an ER network with $N=10^5$ nodes and mean node 
    degree $c=3$. At each time interval $\delta t=1/N$ of the targeted attack
    process, a node chosen by the CI algorithm or by the BPD algorithm is
    deleted along with all the attached links. The three sets of simulation
    data obtained by the CI algorithm correspond to ball radius $\ell=2$ (dotted
    line), $\ell=3$ (dashed line), and $\ell=4$ (long-dashed line), 
    respectively.
    The BPD results (solid line) are obtained at fixed re-weighting parameter
    $x=12$.
  }
\end{figure}

As an example we plot in Fig.~\ref{fig:ERattack} the relative size $g(t)$ of
the largest connected component of an ER network with mean node degree $c=3$.
Initially the network has a giant component of relative size
$g(0) \approx 0.9412$; this giant component then shrinks with time $t$
gradually and finally disappears when about $0.16 N$ nodes are removed.

The results of the CI algorithm are not sensitive to the particular choice of
the ball radius $\ell$ (see Fig.~\ref{fig:ERattack} and discussions in
\cite{Morone-Makse-2015}). For simplicity we fix $\ell=4$ in the remaining part
of this paper, except for the two smallest networks of Table~\ref{tab:realnet} 
(for which $\ell=2$ is used).
To decrease the algorithm's time complexity, in each decimation step a tiny
fraction $f$ of the nodes (instead of a single node) is deleted from the
network and then the CI strengths of the remaining nodes are updated. The 
precise value of $f$ does not affect the quality of the final solution as
long as $f$ is sufficiently small (e.g., $f=0.001$) \cite{Morone-Makse-2015}.
The authors of \cite{Morone-Makse-2015} have made
the source code of the CI algorithm publically available through their webpage.
We use their code in the present comparative study.

\subsection{Belief propagation-guided decimation}
\label{subsec:BPD}

\begin{figure}
  \begin{center}
    \includegraphics[angle=270,width=0.45\textwidth]{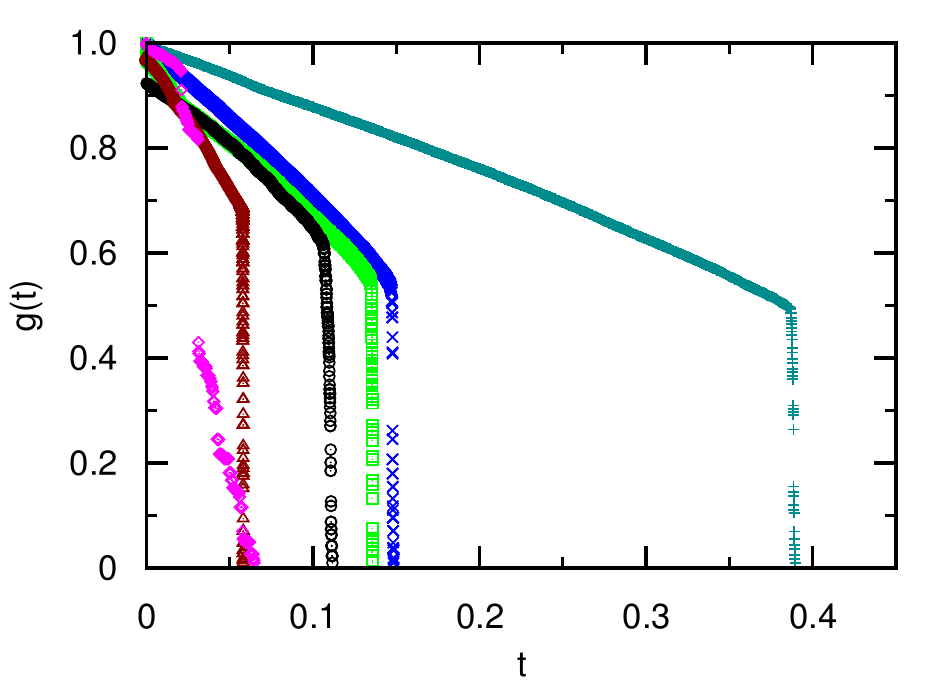}
  \end{center}
  \caption{\label{fig:realnet}
    The relative size $g(t)$ of the largest connected component at algorithmic
    time $t$ of the BPD-guided attack process, for six real-world networks of
    different sizes $N$ (see Table~\ref{tab:realnet}): Citation (pluses),
    P2P (crosses), Friend (squares), Authors (circles), WebPage (triangles),
    Grid (diamonds). At each time interval $\delta t=1/N$ of the targeted
    attack process, a node chosen by the BPD algorithm (with $x=12$) is deleted
    along with all the  attached links.
  }
\end{figure}

The BPD algorithm is rooted in the spin glass model for the feedback vertex
set problem \cite{Zhou-2013}. At each time point $t$ of the iteration process,
the algorithm estimates the probability $q_i^0(t)$ that every node $i$ of the
remaining network $G(t)$ is suitable to be deleted. The explicit formula for
this probability is
\begin{equation}
  \label{eq:qi0}
  q_i^0 = \frac{1}{1+ e^{x} \Bigl[1+ \sum\limits_{k\in \partial i(t)}
      \frac{(1-q_{k\rightarrow i}^0)}{q_{k\rightarrow i}^0 + q_{k\rightarrow i}^k}
      \Bigr] \prod\limits_{j\in \partial i(t)}
    [q_{j\rightarrow i}^{0}+q_{j\rightarrow i}^j] } \; ,
\end{equation}
where $x$ is an adjustable re-weighting parameter, and $\partial i(t)$ denotes
node $i$'s set of neighboring nodes at time $t$. The quantity
$q_{j\rightarrow i}^0(t)$ in Eq.~(\ref{eq:qi0}) is the probability that the
neighboring node $j$ is suitable to be deleted if node $i$ is absent from the
network $G(t)$, while $q_{j\rightarrow i}^j(t)$ is the probability that this node
$j$ is suitable to be the root node of a tree component in the absence of node
$i$ \cite{Zhou-2013}. These two auxiliary probability values are estimated
self-consistently through the following set of belief propagation (BP)
equations:
\begin{subequations}
  \label{eq:BP}
  \begin{align}
    q_{i\to j}^0 & = \frac{1}{z_{i \to j}(t)} \; , \\
    q_{i\to j}^i & = \frac{e^{x} \prod_{k \in \partial i(t) \backslash j}
      \bigl[ q_{k \to i}^0 + q_{k \to i}^k \bigr]}{z_{i \to j}(t)} \; ,
  \end{align}
\end{subequations}
where $\partial i(t) \backslash j$ is the node subset obtained by removing node
$j$ from set $\partial i(t)$, and $z_{i\rightarrow j}(t)$ is a normalization
constant determined by
\begin{eqnarray}
  z_{i \rightarrow j}(t) & = & 1+ e^{x} \prod\limits_{k\in \partial i(t) \backslash j}
  \bigl[ q_{k \to i}^0 + q_{k \to i}^k \bigr] \nonumber \\
  & & \quad \quad \times \Bigl[1+ \sum\limits_{l \in \partial i(t) \setminus j}
    \frac{(1-q_{l \to i}^0)}{q_{l \to i}^0 + q_{l \to i}^l} \Bigr] \; .
  \label{eq:zij}
\end{eqnarray}

At each time step $t$, we first iterate the BP equation (\ref{eq:BP}) on the
network $G(t)$ a number of rounds, and then use Eq.~(\ref{eq:qi0}) to estimate
the probability of choosing each node $i\in G(t)$ for deletion. The node with
the highest probability of being suitable for deletion is deleted from network
$G(t)$ along with all its attached links. The algorithmic time then increases
to $t\leftarrow t+ \frac{1}{N}$ and the next BPD iteration begins. This node
deletion process stops after all the loops in the network have been destroyed
\cite{Zhou-2013}. Then we check the size of each tree component in the
remaining network. If a tree component is too large (which occurs only rarely),
we delete an appropriately chosen node from this tree to achieve a maximal
decrease in the tree size (see Appendix~\ref{sec:appendix} for details). 
We repeat this node deletion process until all the tree components are
sufficiently small.

As an illustration of the BPD iteration process, we record in
Fig.~\ref{fig:ERattack} (solid line) the relative size $g(t)$ of the largest
connected component of an ER random network at each algorithmic time $t$. At
$t\approx 0.137$ the BPD-guided attack stops, resulting in a final target node
set of size $\approx 0.137 N$. Qualitatively similar plots are obtained for
real-world network instances (see Fig.~\ref{fig:realnet}).

\begin{figure}[h]
  \begin{center}
    \includegraphics[angle=270,width=0.45\textwidth]{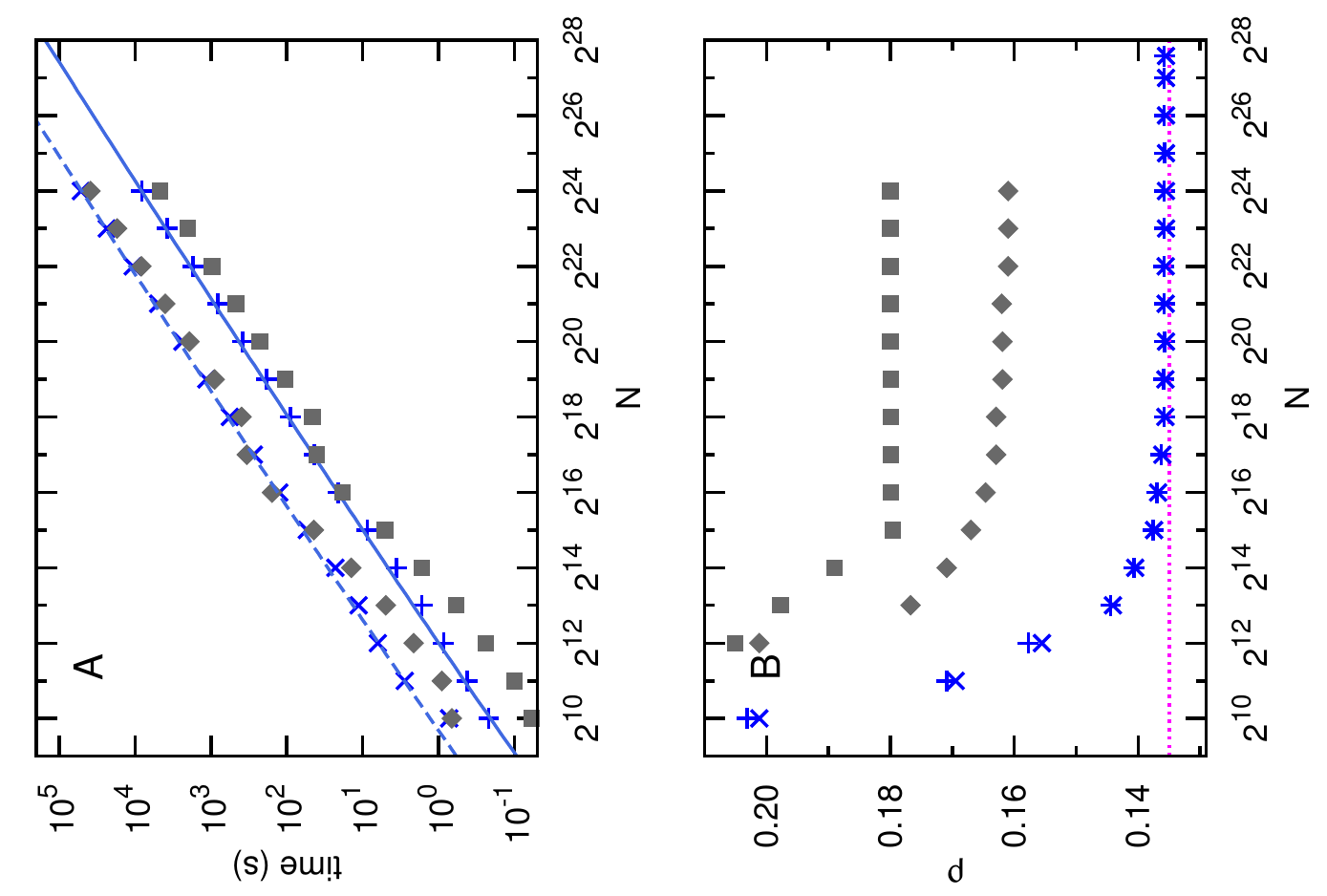}
  \end{center}
  \caption{\label{fig:BPDtimecomp}
    Performance of the BPD algorithm ($x=12$) and the CI algorithm ($\ell=4$)
    on ER networks of mean degree $c=3$ and size $N$. The results
    obtained by BPD are shown as plus symbols (for decimation
    fraction $f=0.01$) and cross symbols (for $f=0.001$), while the results
    obtained by CI are shown as square symbols (for  $f=0.01$) and
    diamond symbols (for $f=0.001$).
    (A) The relationship between the total running time and
    $N$. The simulation results are obtained on a relatively
    old desktop computer (Intel-6300, $1.86$ GHz, $2$ GB memory). The solid
    and dashed lines are the fitting curves of the form 
    $T_{BPD} = a N \ln (N)$ for the BPD running time $T_{BPD}$, with 
    parameter $a=2.896 \times 10^{-5}$ second (for $f=0.01$)
    and $a=1.814 \times 10^{-4}$ second (for $f=0.001$), respectively. BPD
    CI have comparable time complexity for this graph ensemble.
    (B) The relationship between the relative size $\rho$ of the target node
    set and $N$. 
    The dotted horizontal line denotes the predicted minimum value of
    $\rho= 0.13493$ (for $N=\infty$) 
    by the replica-symmetric mean field theory  \cite{Zhou-2013}.
    The BPD results obtained at decimaion fraction $f=0.01$ are almost equally
    good as those obtained at $f=0.001$, but the CI results at 
    $f=0.01$ are much worse than the CI results obtained at $f=0.001$.
  }
\end{figure}

Similar to the CI algorithm, when the BPD algorithm is used as a heuristic
solver, we delete in each decimation step a tiny fraction $f$ of the nodes
in network $G(t)$ and then update the
probability $q_i^0$ for each remaining node $i$.
The BPD algorithm is very fast. It finishes in few minutes when applied on the
large example network of Fig.~\ref{fig:ERattack} and most of the network
instances of Table~\ref{tab:realnet}. In terms of scaling, if the link number
$M$ of the network is of the same order as the node number $N$ (i.e., the
network is sparse), then the running time of the BPD algorithm is proportional
to $N \ln N$ (see Fig.~\ref{fig:BPDtimecomp}A for a concrete demonstration).
This algorithm therefore is applicable to extremely huge network instances.
For example, when applied on an ER network with $N=2 \times 10^8$ nodes
and mean degree $c=3$, the BPD algorithm returns a target node set of
relative size $\rho=0.13574$ in $23.50$ hours ($f=0.01$) and another
solution of relative size $\rho=0.13610$ in $14.68$ hours ($f=0.03$). These
relative sizes are just $0.6\%$ beyond the predicted 
minimum relative size of $\rho=0.13493$ by the replica-symmetric mean field
theory \cite{Zhou-2013}. In contrast, the relative sizes of the solutions
obtained by 
the CI algorithm are $23.7\%$ (for $\ell=2$), $21.5\%$
($\ell=3$), and $20.8\%$ ($\ell=4$) beyond the prediction of the
replica-symmetric mean field theory.

The original BPD code for the minimum feedback vertex set problem and its
slightly adjusted version for the network optimal attack problem are both
available at
{\tt power.itp.ac.cn/$\sim$zhouhj/codes.html}.
The BPD results of the next section are obtained at fixed value of 
deletion fraction $f=0.01$. As the example results of 
Fig.~\ref{fig:BPDtimecomp}B further demonstrate, the precise value of $f$
does not affect the relative size $\rho$ of the BPD solutions.

\subsection{Gradual decrease versus abrupt drop}

Figure~\ref{fig:ERattack} clearly shows that, compared to the CI algorithm, the
BPD algorithm constructes a much smaller target node set for the same ER
network instance. This superiority holds true for all the networks we examined
(see next section). We also notice from Fig.~\ref{fig:ERattack} that, during
the CI-guided attack process the size of the giant connected component
decreases gradually and smoothly. On the other hand, if the attacked nodes are
chosen according to the BPD algorithm, the giant component initially shrinks
slowly and almost linearly and the decrease in size is roughly equal to the
increase in the number of deleted nodes; but as the giant component's relative
size reduces to $\approx 0.76$ after a fraction $\approx 0.133$ of the nodes
are deleted, the network is in a very fragile state and the giant component
suddenly disappears with the deletion of an additional tiny fraction of nodes.

Such an abrupt collapse phenomenon, which resembles the phenomenon of explosive
percolation
\cite{Achlioptas-DSouza-Spencer-2009,Riordan-Warnke-2011,Cho-etal-2013},
is also observed in the BPD-guided attack processes on other random network
ensembles and real-world networks (Fig.~\ref{fig:realnet}). It may be a generic
feature of the BPD-guided network attack. Indeed the BPD algorithm is not
designed to break a network down into small pieces but is designed to cut loops
in the most efficient way. This loop-cutting algorithmic design principle may
explain why the collapse of a giant connected component occurs at the latest
stage of the attack process and is abrupt. We expect that during the BPD-guided
attack process, the most significant changes in the network is that the number
of loops in the giant components decreases quickly. A highly connected node
that bridges two or more parts of the network will only have a low probability
of being deleted if it does not contribute much to the loops of the network
\cite{Zhou-2013}.

\section{Comparative results}
\label{sec:results}

We now apply the CI and the BPD algorithm on a large number of network
instances. We adopt the same criterion used in \cite{Morone-Makse-2015}, namely
that after the deletion of a set $S$ of nodes the largest connected component
should have relative size $\leq \theta=0.01$. The size of this set $S$
(relative to the total number $N$ of vertices) is denoted as $\rho$, see
Fig.~\ref{fig:BPDtimecomp}B, Fig.~\ref{fig:ERandRR},
and Fig.~\ref{fig:SFandSF}.

Following \cite{Morone-Makse-2015}, when applying the CI algorithm to a network
$G$, we first delete a draft set of nodes from the network until the largest
component of the remaining network contains no more than $\theta N$ nodes. We
then refine this set by sequentially moving some nodes back to the network.
Each of these displaced nodes has the property that its addition to the network
will not cause an increase in the size of the largest network component and
will only cause the least increase in the size of a small component. The final
set $S$ of deleted nodes after this refinement process is regarded as a
solution to the optimal network attack problem. This same refinement process is
also adopted by the BPD algorithm. We first apply BPD to construct a FVS for
the input network, then a few additional nodes are deleted break very large
trees. Finally some of the nodes in the deleted node set $S$ are added back to
the network as long as they cause the least perturbation to the largest
connected component and its increased relative size is still below $\theta$.
This refinement process recovers some of the deleted short loops.

\subsection{ER and RR network ensembles}

We first consider Erd\"os-Reny\'i random networks and regular random networks.
An ER network of $N$ vertices and $M = (c/2)N$ links is generated  by first
selecting $M$ different node pairs uniformly at random from the whole set of
$N (N -1)/2$ candidate  pairs and then add a link between the chosen two nodes.
Each node in the network has $c$ attached links on average. A RR network is
more regular in the sense that each node has exactly the same number $K$ of
nearest neighbors; it is generated by first attaching to each node $K$
half-links and then randomly connecting two half-links into a full link
(excluding self-links and multiple-links).

\begin{figure}
  \begin{center}
    \includegraphics[angle=270,width=0.45\textwidth]{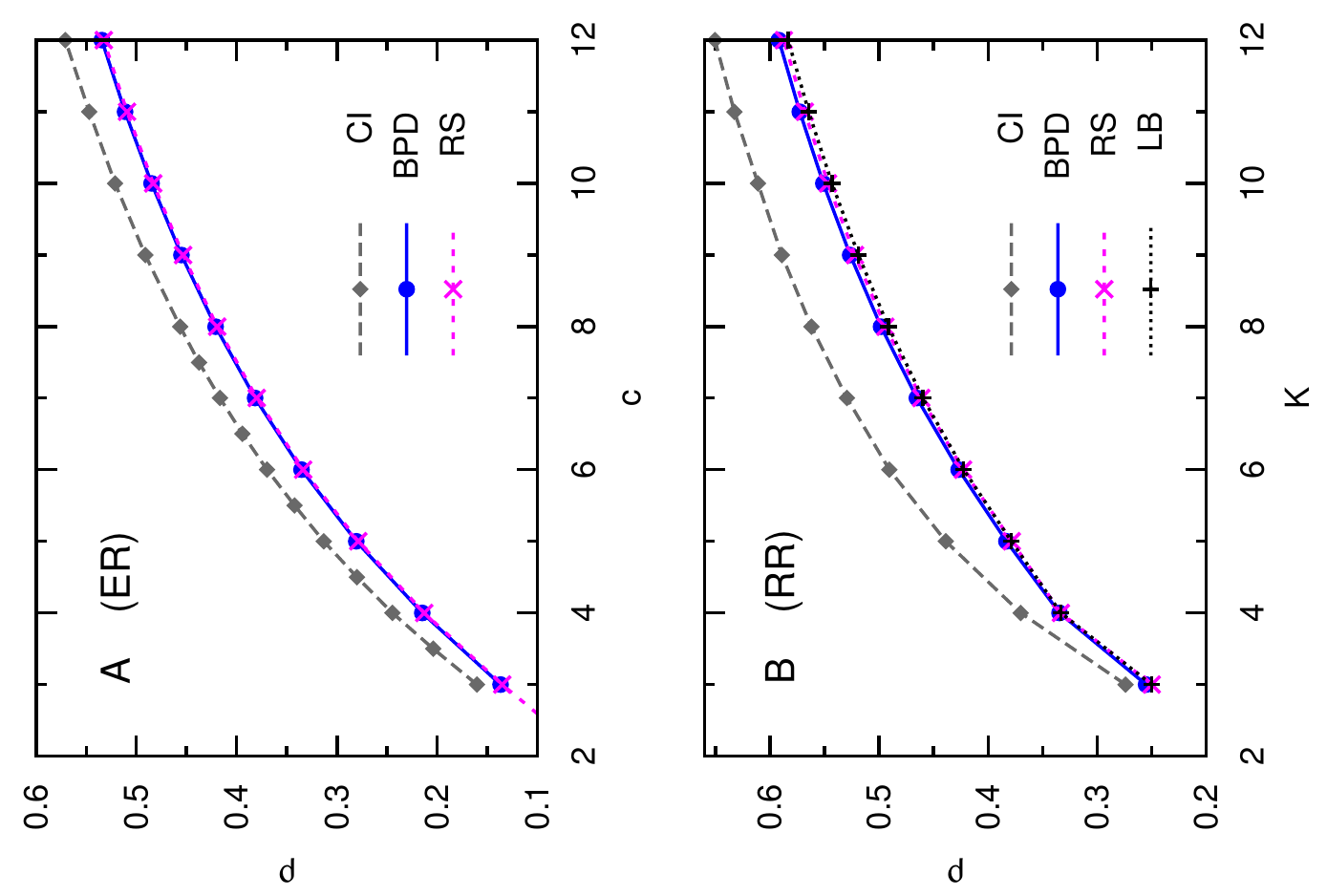}
  \end{center}
  \caption{  \label{fig:ERandRR}
    Fraction $\rho$ of removed nodes for Erd\"{o}s-R\'enyi random networks of
    mean degree $c$ (A) and regular random networks of degree $K$ (B). Each CI
    (diamond) or BPD (circle) data point is the averaged result over $48$
    network instances (size $N=10^5$); the standard deviation is of order
    $10^{-4}$ and is therefore not shown. The cross symbols are the predictions
    of the replica-symmetric (RS) mean field theory on the minimum relative 
    size of the target node sets \cite{Zhou-2013}. The plus symbols of (B) are
    the mathematical lower bound (LB) on the minimum relative size of the
    target node sets \cite{Bau-Wormald-Zhou-2002}. The re-weighting parameter
    of the BPD algorithm is fixed to $x= 12$ for ER networks and $x=7$ for RR
    networks; the ball radius parameter of the CI algorithm is fixed to
    $\ell=4$.
  }
\end{figure}

The target node set $S$ for breaking down a random network contains an
extensive number $\rho N$ of nodes. We find that the BPD algorithm obtains 
qualitatively better solutions than the CI algorithm,  in the sense that 
$\rho_{BPD}$ is much smaller than $\rho_{CI}$ (Fig.~\ref{fig:ERandRR}). For
example, the CI-guided attack scheme would need to delete a fraction
$\rho_{CI}\approx 0.52$ of all the nodes to break down an ER network of mean
degree $c=10$, while the BPD-guided scheme only needs to delete a smaller
fraction $\rho_{BPD}\approx 0.48$. The difference in performance between CI and
BPD is even more pronounced on RR networks (Fig.~\ref{fig:ERandRR}B).

Indeed there is not much room to further improve over the BPD algorithm. As we
show in Fig.~\ref{fig:ERandRR} the value of $\rho_{BPD}$ almost overlaps with
the predicted minimum value by the replica-symmetric mean field (which is
non-rigorously believed to be a lower bound to the true minimum value). For the
RR network ensemble, the value of $\rho_{BPD}$ is also very close to the
rigorously known lower bound for the minimum value
\cite{Bau-Wormald-Zhou-2002}, while the empirical value $\rho_{CI}$ obtained by
the CI algorithm is far beyond this mathematical bound
(Fig.~\ref{fig:ERandRR}B).

\subsection{Scale-free random network ensembles}
\label{subsec:SF}

We then examine random scale-free (SF) networks. The static method
\cite{Goh-Kahng-Kim-2001} is followed to generate a single SF network instance.
Each node $i \in \{1, 2, \ldots, N\}$  is assigned a fitness value
$f_{i}=i^{-\xi}/\sum_{j=1}^{N} j^{-\xi}$ with $0 \leq  \xi < 1$ being a fixed
parameter. A total number of $M= (c/2) N$  links are then sequentially added to
the network:  first a pair of nodes $(i, j)$ is chosen from the network with
probability $f_i f_j$ and then a link is added between $i$ and $j$ if it does
not result in a self-link or a multiple-link. The resulting network has a
power-law degree distribution, so the probability of a randomly chosen node to
have $d \gg 1 $ attached links is proportional to $d^{-\gamma}$ with the decay
exponent being $\gamma=1+1/ \xi$  \cite{Goh-Kahng-Kim-2001}. There are many
highly connected (hub) nodes in a SF network, the degrees of which greatly
exceed the mean node degree $c$. 

\begin{figure}
  \begin{center}
    \includegraphics[angle=270,width=0.45\textwidth]{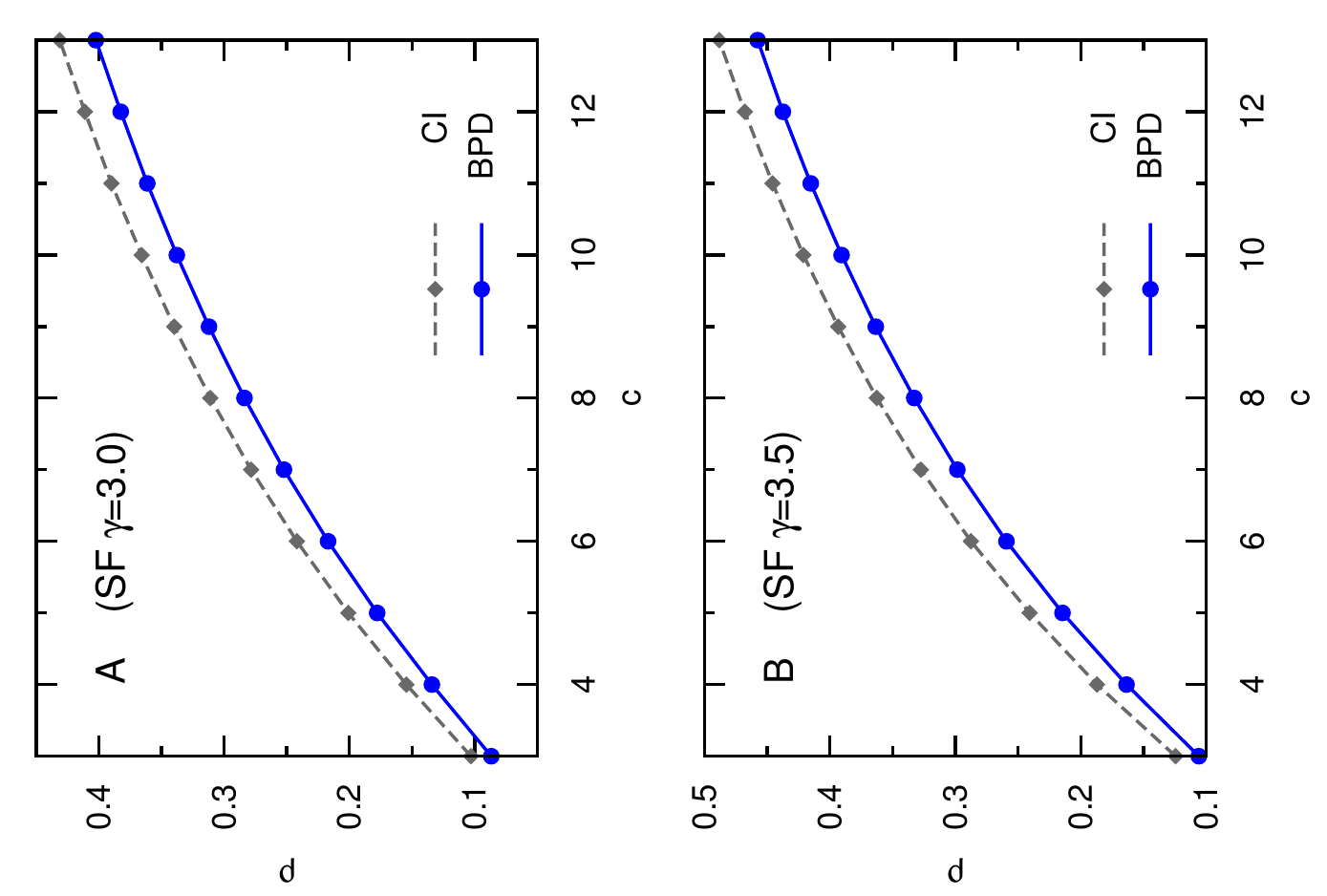}
  \end{center}
  \caption{\label{fig:SFandSF}
    Fraction $\rho$ of removed nodes for scale-free random networks of mean
    degree $c$ and degree decay exponent $\gamma = 3.0$ (A) and $\gamma=3.5$
    (B). Each CI (diamond) or BPD (circle) data point is the averaged result
    over $48$ network instances (size $N=10^5$) generated through the static
    method \cite{Goh-Kahng-Kim-2001}; the standard deviation (not shown) of
    each data point is of order $10^{-4}$. The re-weighting parameter of the
    BPD algorithm is fixed to $x= 12$; the ball radius parameter of the CI
    algorithm is fixed to $\ell=4$.
  }
\end{figure}

The BPD-guided attack scheme is again qualitatively more efficient than the
CI-guided attack scheme (Fig.~\ref{fig:SFandSF}). For example, the BPD
algorithm only needs to delete a fraction $\rho_{BPD}\approx 0.338$ of all the
nodes to break down a SF network with mean degree $c=10$ and decay exponent
$\lambda=3.0$, while the CI algorithm would need to delete a larger fraction
$\rho_{CI}\approx 0.366$ of the nodes. We have also considered random SF
networks with decay exponent $\gamma=2.67$ and $\gamma=2.5$. The obtained
results are qualitatively the same as those shown in Fig.~\ref{fig:SFandSF}.
At the same mean node degree $c$, the gap between $\rho_{CI}$ and $\rho_{BPD}$
seems to enlarge slowly with the power-law exponent $\gamma$.

Since there exist many hub nodes, one would expect that the optimal attack
problem is easier to solve on SF random networks than on homogeneous network.
Seeing that the BPD algorithm performs perfectly for ER and RR random networks,
we anticipate that the solutions obtained on SF networks are also very close to
be minimum targeted attack sets. Further computer simulations
\cite{Qin-Zhou-2014,Galinier-Lemamou-Bouzidi-2013} and replica-symmetric mean
field computations \cite{Zhou-2013} need to be carried out to confirm this
conjecture.

\subsection{Real World Network}
\label{sec:real}

\begin{table}[t]
  \caption{
    \label{tab:realnet}
    Comparative results of the CI and the BPD algorithm on a set of real-world
    network instances. $N$ and $M$ are the number of nodes and links of each
    network, respectively. The targeted attack set (TAS) sizes obtained by CI
    and BPD are listed in the 4th and 5th column, and the feedback vertex set
    (FVS) sizes obtained by these algorithms are listed in the 6th and 7th
    column. The BPD algorithm is run with fixed re-weighting parameter $x=12$,
    and the ball radius parameter of CI is fixed to $\ell=4$ except for the
    RoadEU and the PPI network, for which $\ell=2$.
  }
  \begin{center}
    \begin{footnotesize}
      \begin{tabular}{|l|r|r|rr|rr|}
        \hline \hline
        \multirow{2}{*}{Network}  &\multirow{2}{*}{$N$}  &
        \multirow{2}{*}{$M$}         & \multicolumn{2}{c|}{TAS}  &
        \multicolumn{2}{c|}{FVS} \\
        \cline{4-7}
        & & & CI & BPD & CI & BPD \\
        \hline 
        RoadEU   &    $1177$ &     $1417$ & $209$         & $152$ 
        & $107$ & $91$ \\
        PPI      &    $2361$ &     $6646$ & $424$         & $350$
        & $391$ & $362$ \\
        Grid     &    $4941$ &     $6594$ & $476$         & $320$
        & $663$ & $512$ \\
        IntNet1  &    $6474$ &    $12572$ & $198$         & $161$
        & $248$ & $215$ \\ 
        Authors  &   $23133$ &    $93439$ & $3588$   &  $2583$
        & $9429$ & $8317$ \\
        Citation &   $34546$ &   $420877$ & $14518$       & $13454$
        & $16470$ & $15390$ \\
        P2P      &   $62586$ &   $147892$ & $10726$       & $9292$ 
        & $9710$ & $9285$ \\
        Friend   &  $196591$ &   $950327$ & $32340$   & $26696$
        & $48425$ & $38831$ \\
        Email    &  $265214$ &   $364481$ & $21465$       & $1064$
        & $20801$ & $1186$ \\
        WebPage  &  $875713$ &  $4322051$ & $106750$      & $50878$
        & $257047$ & $208641$ \\
        RoadTX   & $1379917$ &  $1921660$ & $133763$ &  $20676$
        & $319128$ & $239885$ \\
        IntNet2  & $1696415$ & $11095298$ & $144160$  &  $73229$
        & $318447$  & $228720$ \\
        \hline \hline
      \end{tabular} 
    \end{footnotesize}
  \end{center}
\end{table}  

Finally we compare CI and BPD on real-world network instances, which are
usually not completely random nor completely regular but have rich local and
global structures (such as communities and hierarchical levels).
Table~\ref{tab:realnet} lists the twelve network instances considered in this
work. There are five infrastructure networks: the European express road network
(RoadEU \cite{Subelj-Bajec-2011}), the road network of Texas  (RoadTX
\cite{Leskovec-etal-2009}), the power grid of western US states (Grid
\cite{Watts-Strogatz-1998}), and two Internet networks at the autonomous
systems level (IntNet1 and IntNet2 \cite{Leskovec-etal-2005}). Three of the
remaining networks are information communication networks: the Google webpage
network (WebPage \cite{Leskovec-etal-2009}), the European email network (Email
\cite{Leskovec-Kleinberg-Faloutsos-2007}), and a research citation network
(Citation \cite{Leskovec-etal-2005}). This set also includes one biological
network (the protein-protein interaction network \cite{Bu-etal-2003}) and three
social contact networks: the collaboration network of condensed-matter authors
(Author \cite{Leskovec-Kleinberg-Faloutsos-2007}), a peer-to-peer interaction
network (P2P \cite{Ripeanu-etal-2002}), and an online friendship network
(Friend \cite{Cho-etal-2011}). There are an abundant number of triangles (i.e.,
loops of length three) in these real-world network instances, making the
clustering coefficients of these networks to be considerably large.

For each of these network instances the BPD algorithm constructs a much
smaller targeted attack node set than the CI algorithm does. In some of the
network instances the differences are indeed very remarkable (e.g., the Grid
network, the Email network and the RoadTX network in Table~\ref{tab:realnet}).
When we compare the sizes of the feedback vertex sets we again observe
considerable improvements of the BPD algorithm as compared to the CI algorithm.

Similar to what happens on random networks (Fig.~\ref{fig:ERattack}), when the
BPD-guided attack scheme is applied to these real-world networks, the giant
network components do not change gradually but experience abrupt collapse
transitions (see Fig.~\ref{fig:realnet} for some examples).

\section{Conclusion and discussions}
\label{sec:conclude}

In this work we demonstrated that the network optimal attack problem, a central
and difficult optimization problem in network science, can be solved very
efficiently by a BPD message-passing algorithm that was originally proposed to
tackle the network feedback vertex set problem \cite{Zhou-2013}. In terms of
time complexity, the BPD algorithm is almost a linear algorithm (see
Fig.~\ref{fig:BPDtimecomp}A), so it is applicable even to extremely huge
real-world networks. Our numerical results also demonstrated that the local
Collective Information algorithm of \cite{Morone-Makse-2015} can not offer
nearly optimal solutions to the network optimal attack problem (which was
re-named as the network optimal influence problem in \cite{Morone-Makse-2015}).
As an empirical algorithm designed to cut loops most efficiently, the BPD will
be very useful in network resilience studies and in help identifying the most
influential nodes.

Another major observation was that the BPD-guided attach causes an abrupt
breakdown of the network. This latter dynamical property, combined with
requiring only a minimum number of target nodes, may make the BPD-guided
attack scheme a dangerous strategy if it is adopted for destructive purposes.
The society might need to seriously evaluate such a potential threat and, if
necessary, to implement suitable prevention protocols. An anoymous reviewer
suggested to us that it might be sufficient to compare the largest eigenvalue
of the network's non-backtracking matrix (also called the Hashimoto matrix)
\cite{Krzakala-etal-2013,Hashimoto-1989} to distinguish between an intentional
attack and a randomized node deletion process. We hope to explore this
interesting idea in a separate paper.

For simplicity we assumed in this paper that the cost $w_i$  of deleting a node
$i$ is the same for different nodes, i.e., $w_i = 1$ for $i=1, 2, \ldots, N$.
Let us emphasize that if this cost is not uniform but is node-specific, the BPD
algorithm is also appicable \cite{Zhou-2013}. The only essential modification
is that the re-weighting factor $e^{x}$ in Eqs.~(\ref{eq:qi0}), (\ref{eq:BP})
and (\ref{eq:zij}) should be replaced by $e^{x w_i}$.

\emph{Note Added:} Several closely related papers appeared on the arXiv e-print
server after the first version of this manuscript was posted on arXiv and
submitted for review. The paper of Braunstein and co-authors
\cite{Braunstein-etal-2016} considered the network optimal attack problem also
as a minimum feedback vertex set problem, while the paper of Clusella and
co-authors \cite{Clusella-etal-2016} applied the idea of explosive percolation
to the optimal attack problem. The paper of Morone and co-authors
\cite{Morone-etal-2016} presented a new version of the CI algorithm which, at
the cost of much increased computing time, may achieve better solutions than
the original CI algorithm.

\section*{Acknowledgments}

HJZ thanks Dr. Yuliang Jin for bringing Ref.~\cite{Morone-Makse-2015} to his
notice. This work was supported by the National Basic Research Program of China
(grant number 2013CB932804), by the National Natural Science Foundation of
China (grant numbers 11121403 and 11225526), and by the Knowledge Innovation
Program of Chinese Academy of Sciences (No.~KJCX2-EW-J02). The first author
(S.M.) is supported by a  CAS-TWAS president fellowship.

\begin{appendix}
  \section{Optimally attacking a tree}
  \label{sec:appendix}
  
  Given a tree $T$ formed by $n$ nodes and $(n-1)$ links, how to choose an
  optimal node  so that after deleting this node the size of the \emph{largest}
  component of the resulting forest achieves the minimum value among all the
  $n$ possible choices of the deleted node? We have implemented a simple
  iterative process to solve this choice problem most efficiently. Here we
  briefly describe this process.
  
  First we consider all the leaves (i.e., nodes of degree one) of tree $T$. For
  each leaf node (say $i$) we know that its deletion will lead to a sub-tree of
  size $(n-1)$, and we let this leaf node to send a message
  $m_{i\rightarrow j}=1$ to its unique neighbor $j$ in tree $T$. After all these
  leaf nodes are considered, we delete them from tree $T$ to obtain a reduced
  tree $T^\prime$, and then we consider all the leaf nodes of $T^\prime$. For
  each leaf node $j\in T^\prime$, we let it to send a message
  $m_{j\rightarrow k}\equiv 1+\sum_{i\in \partial j\backslash k} m_{i\rightarrow j}$ to
  its unique neighbor $k$ in tree $T^\prime$, where $\partial j$ denotes the set
  formed by all the neighboring nodes of $j$ in the \emph{original} tree $T$.
  If node $j$ is deleted from tree $T$, the component sizes of the resulting
  forest will form the following merged set
  $\{m_{l\rightarrow j}| l\in\partial j\backslash k\}
  \cup\{(n-m_{j\rightarrow k})\}$, and we can easily identify the largest member
  of this integer set. After all the leaves of $T^\prime$ have been examined,
  we again delete them to get a further reduced tree $T^{\prime\prime}$ and we
  repeat to consider all the leaf nodes of $T^{\prime\prime}$ in the same way.
  After a few iterations all the nodes in the original tree $T$ will be
  exhausted, and we will be able to identify the optimal node $i$ for breaking
  this tree $T$.
\end{appendix}


\end{document}